\documentclass[sigconf,natbib,9pt]{acmart}

\usepackage{booktabs} 
\usepackage{graphicx}
\usepackage{url}
\usepackage{microtype}
\usepackage{tabularx,amsmath}
\usepackage{amssymb}
\usepackage{multirow}
\usepackage{color}
\usepackage[normalem]{ulem}
\usepackage[english]{babel}
\usepackage{colortbl}
\usepackage[noend]{algorithmic}
\usepackage{algorithm}
\usepackage{paralist}
\usepackage{enumerate}
\usepackage{enumitem}
\usepackage{acronym}
\usepackage[skip=0pt]{caption}
\usepackage{setspace}
\usepackage{balance}

\newbox\itembox
\def\itemlistlabel#1{#1\hfill}
\def\itemlist#1{\setbox\itembox=\hbox{#1}%
                \list{}{\labelwidth\wd\itembox
                             \leftmargin\labelwidth
                             \advance\leftmargin by\itemindent
                             \advance\leftmargin by\labelsep
                             \let\makelabel\itemlistlabel}}

\newcommand{\enkelop}{$^{\vartriangle}$}
\newcommand{\dubbelop}{$^{\blacktriangle}$}
\newcommand{\enkelneer}{$^{\triangledown}$}
\newcommand{\dubbelneer}{$^{\blacktriangledown}$}

\acrodef{IR}{Information Retrieval}
\acrodef{LTR}{Learning to Rank}
\acrodef{OLTR}{Online Learning to Rank}
\acrodef{MGD}{Multileave Gradient Descent}
\acrodef{C-MGD}{Cascading Multileave Gradient Descent}
\acrodef{Sim-MGD}{Similarity Model}
\acrodef{RL}{Reinforcement Learning}

\newif\ifanon
\ifx\anonymous\undefined
  \anonfalse
\else
  \anontrue
\fi

\title[Balancing Speed and Quality]{Balancing Speed and Quality in Online Learning to Rank for Information Retrieval}

\author{Harrie Oosterhuis}
\orcid{}
\affiliation{%
\institution{University of Amsterdam}
\city{Amsterdam}
\country{The Netherlands}
}
\email{oosterhuis@uva.nl}

\author{Maarten de Rijke}
\orcid{0000-0002-1086-0202}
\affiliation{%
\institution{University of Amsterdam}
\city{Amsterdam}
\country{The Netherlands}
}
\email{derijke@uva.nl}

\begin{document}

\begin{abstract}
In \ac{OLTR} the aim is to find an optimal ranking model by interacting with users. 
When learning from user behavior, systems must interact with users while simultaneously learning from those interactions. Unlike other \ac{LTR} settings, existing research in this field has been limited to linear models. This is due to the speed-quality tradeoff that arises when selecting models: complex models are more expressive and can find the best rankings but need more user interactions to do so, a requirement that risks frustrating users during training. Conversely, simpler models can be optimized on fewer interactions and thus provide a better user experience, but they will converge towards suboptimal rankings. This tradeoff creates a deadlock, since novel models will not be able to improve either the user experience or the final convergence point, without sacrificing the other.

Our contribution is twofold. First, we introduce a fast \ac{OLTR} model called Sim-MGD that addresses the speed aspect of the speed-quality tradeoff. Sim-MGD ranks documents based on similarities with reference documents. It converges rapidly and, hence, gives a better user experience but it does not converge towards the optimal rankings.
Second, we contribute \ac{C-MGD} for \acs{OLTR} that directly addresses the speed-quality tradeoff by using a cascade that enables combinations of the best of two worlds: fast learning and high quality final convergence. \ac{C-MGD} can provide the better user experience of Sim-MGD while maintaining the same convergence as the state-of-the-art \acs{MGD} model. This opens the door for future work to design new models for \acs{OLTR} without having to deal with the speed-quality tradeoff.
\end{abstract}

%

\copyrightyear{2017} 
\acmYear{2017} 
\setcopyright{acmlicensed}
  \acmConference{CIKM'17}{}{November 6--10, 2017, Singapore.}
  \acmPrice{15.00}
  \acmDOI{https://doi.org/10.1145/3132847.3132896}
  \acmISBN{ISBN 978-1-4503-4918-5/17/11}
\maketitle


\section{Introduction}
\label{sec:intro}

The goal of \acf{LTR} in \acf{IR} is to optimize models that rank documents according to user preferences. As modern search engines may combine hundreds of ranking signals they rely on models that can combine such signals to form optimal rankings. Traditionally, this was done through Offline Learning to Rank, which relies on annotated sets of queries and documents with their relevance assessed by human raters. Over the years, the limitations of this supervised approach have become apparent: annotated sets are expensive and time-consuming to produce \cite{letor,Chapelle2011}; in some settings creating such a dataset would be a serious breach of privacy \cite{najork2016using, wang2016learning}; and annotations are not necessarily in line with user preferences \cite{sanderson2010}. As a reaction, interest in \acf{OLTR}, where models learn from interactions with users,  has increased \cite{chaudhuri2016online, zhao2016constructing, schuth2016mgd, oosterhuis2016probabilistic}. While this resolves many of the issues with the offline \ac{LTR} setting, it brings challenges of its own. Firstly, \ac{OLTR} algorithms cannot directly observe their performance and thus have to infer from user interactions how they can improve. Secondly, they have to perform their task, i.e., decide what rankings to display, while simultaneously learning from user interactions.

\begin{figure}[h]
\centering
\includegraphics[clip,trim=0mm 2mm 0mm 0mm,width=\columnwidth]{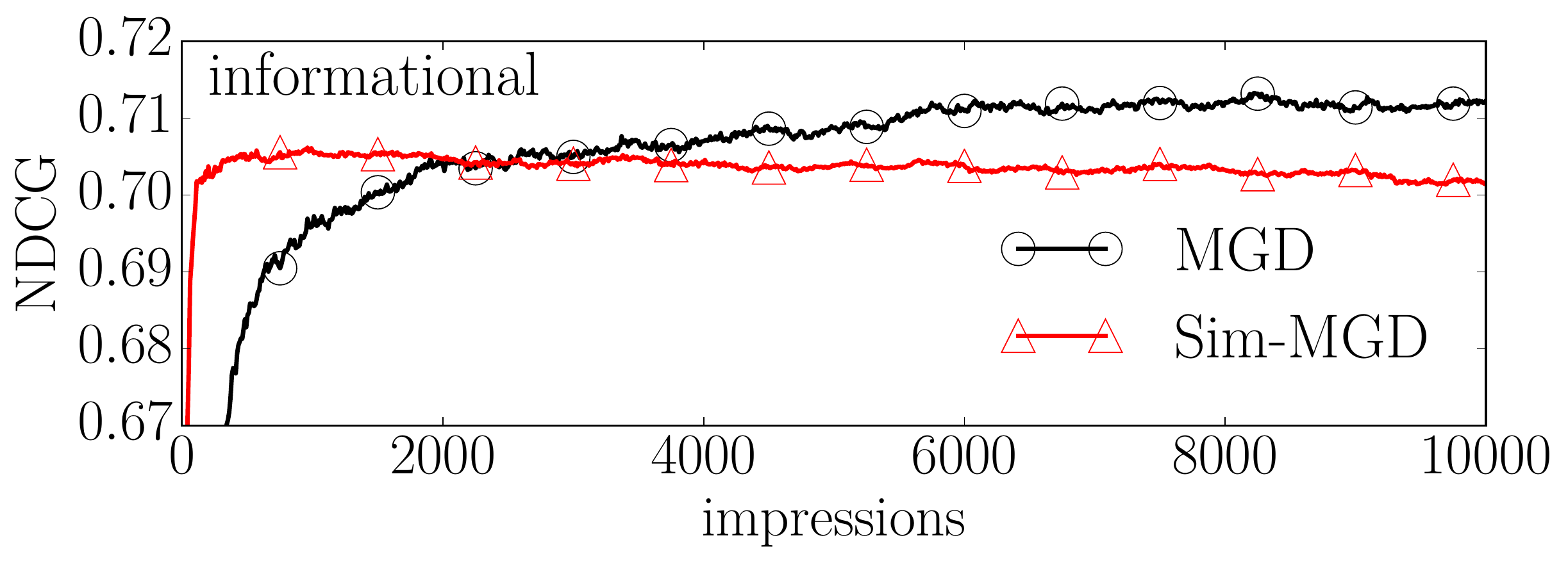}

\caption{Offline performance (NDCG) of a Linear Model (MGD) and a Similarity Model (Sim-MGD) on the NP2003 dataset under an informational click model. (Full details of the experimental setup are provided in Section~\ref{sec:experiments}.)}
\label{fig:intro:tradeoff}
\end{figure}

In stark contrast with other work on \ac{LTR}, existing work in \ac{OLTR} has only considered optimizing linear models and merely focussed on improving gradient estimation. We argue that this limitation is due to a \emph{speed-quality tradeoff} that previous work has faced. This tradeoff is a result of the dual nature of the \ac{OLTR} task: algorithms are evaluated both on how they perform the task while learning and on the final ranking model they converge towards. This duality is especially important as \ac{OLTR} involves human interactions: some strategies may result in an optimal ranking model but may frustrate users during learning.
Consider the experiment visualized in Figure~\ref{fig:intro:tradeoff}. Here, a Linear Model (MGD) and a simpler \ac{Sim-MGD} are optimized on user interactions. The latter learns faster and fully converges in fewer than 200 impressions, while the Linear Model initially trails Sim-MGD but is more expressive, requires more impressions, and ultimately exceeds Sim-MGD in offline performance (as measured in NDCG).

\ac{OLTR} models that are less complex, i.e., that require fewer user interactions to converge, may provide a good user experience as they adapt quickly. However, because of their limited complexity they often lack expressiveness, causing them to learn suboptimal rankings. Conversely, a more complex \ac{OLTR} model may ultimately find the optimal rankings but requires more user interactions. Thus, such models ultimately produce a better experience but risk deterring users before this level of performance is reached. As a result, a fundamental tradeoff has to be made: a good user experience during training resulting in suboptimal rankings vs.\ the risk of frustrating users while finding superior rankings in the end. We call this dichotomy the \emph{speed-quality tradeoff}. 

To address the speed-quality tradeoff, a method for combining the properties of multiple models is required. In this paper we meet this challenge by making two contributions. First, we introduce a novel model that uses document feature similarities (\ac{Sim-MGD}) to learn more rapidly than the state-of-the-art, \ac{MGD} \cite{schuth2016mgd, oosterhuis2016probabilistic}. However, \ac{Sim-MGD} converges towards rankings inferior to \acs{MGD} as predicted by the speed-quality tradeoff.
Secondly, we propose a novel cascading \ac{OLTR} approach, called \ac{C-MGD}, 
that uses two \ac{OLTR} models, a fast simple model and a slower complex model. Initially the cascade lets the faster model learn by interacting with its users. Later, when the faster learner has converged it is used to initialize the expressive model and discarded. \ac{C-MGD} then continues optimization by letting the expressive model interact with the user.
Consequently, the user experience is improved, both short term and long term, as users initially interact with a fast adapting model, while ultimately the better ranker using the complex model is still found.
Our empirical results show that the cascade approach, i.e., \ac{C-MGD}, can combine the improved user experience from \ac{Sim-MGD} while still maintaining the optimal convergence of the state-of-the-art.


In this paper we address the following research questions:
 \begin{enumerate}[label={\bf RQ\arabic*},leftmargin=*,nosep,topsep=1pt]
    \item Is the user experience significantly improved when using \ac{Sim-MGD}? \label{rq:simmgd}
    \item Can the cascading approach, \ac{C-MGD}, combine an improvement in user experience while maintaining convergence towards state-of-the-art performance levels? \label{rq:cmgd}
\end{enumerate}
To facilitate replicability and repeatability of our findings, we provide open source implementations of both \ac{Sim-MGD} and~\ac{C-MGD}.\footnote{\url{https://github.com/HarrieO/BalancingSpeedQualityOLTR}} 




\section{Related Work}
\label{sec:relatedwork}

We provide a brief overview of  \ac{LTR} and \ac{OLTR} before describing methods for combining multiple models in Machine Learning.

\subsection{Learning to rank}
\acf{LTR} is an important part of \acf{IR} and allows modern search engines to base their rankings on hundreds of relevance signals \cite{liu2009learning}. Traditionally, a supervised approach is taken where human raters annotate whether a document is relevant to a query \cite{Qin2013Letor, Chapelle2011}. Additionally, previous research has considered semi-supervised approaches that use unlabeled sample data next to annotated data \cite{szummer11:semi, Joachims2002}. Both supervised and semi-supervised approaches are typically performed offline, meaning that training is performed after annotated data has been collected. When working with previously collected data, the speed-quality tradeoff does not arise, since users are not involved during training. Consequently, complex and expressive models have been very successful in the offline setting \cite{Joachims2002, burges2010ranknet}. 

However, in recent years several issues with training on annotated datasets have been found. Firstly, gathering annotations is time-consuming and costly \cite{letor, Qin2013Letor, Chapelle2011}, making it infeasible for smaller organisations to collect such data. Secondly, for certain search contexts collecting data would be unethical, e.g., in the context of search within personal emails or documents \cite{wang2016learning}. Thirdly, since the datasets are static, they cannot account for future changes in what is considered relevant. Models derived from such datasets are not necessarily aligned with user satisfaction, as annotators may interpret queries differently from actual users \cite{sanderson2010}.

\subsection{Online learning to rank}
\acf{OLTR} attempts to solve the issues with offline annotations by directly learning from user interactions~\cite{yue09:inter}, as direct interactions with users are expected to be more representative of their preferences than offline annotations~\cite{radlinski08:how}. The task of \ac{OLTR} algorithms is two-fold: they must choose what rankings to display to users while simultaneously learning from interactions with the presented rankings. Although the \ac{OLTR} task can be modeled as a \ac{RL} problem~\cite{sutton1998:introduction}, it differs from a typical \ac{RL} setting because there is no observable reward. 
The main difficulties with performing both aspects of the \ac{OLTR} task come in the form of \emph{bias} and \emph{noise}. Noise occurs when the user's interactions do not represent their true preferences, e.g., users often click on a document for unexpected reasons \cite{sanderson2010}. Bias arises in different ways, e.g., there is selection bias as interactions only involve displayed documents \cite{wang2016learning} and position bias as documents at the top of a ranking are more likely to be considered \cite{yue2010beyond}. These issues complicate relevance inference, since the most clicked documents are not necessarily the most relevant. 

Consequently, state-of-the-art \ac{OLTR} algorithms do not attempt to predict the relevance of single documents. Instead, they approach training as a dueling bandit problem~\cite{yue09:inter} which relies on methods from online evaluation to compare rankers based on user interactions~\cite{radlinski2010comparing, radlinski2013practical}. Interleaving methods combine rankings from two rankers to produce a single result list; from large numbers of clicks on interleavings a preference for one of the two rankers can be inferred~\cite{radlinski2013optimized, hofmann2011probabilistic}. This approach has been extended to find preferences between larger sets of rankers in the form of multileaving~\cite{Schuth2014a,schuth2015probabilistic}. These comparison methods have recently given rise  to \ac{MGD}, a more sensitive \ac{OLTR} algorithm that requires fewer user interactions to reach the same level of performance \cite{schuth2016mgd}. The improvement is achieved by comparing multiple rankers at each user impression, the results of which are then used to update the \ac{OLTR} model. Initially, the number of rankers in the comparison was limited to the SERP length \cite{Schuth2014a}. Probabilistic multileaving~\cite{schuth2015probabilistic} allows comparisons of a virtually unlimited size, leading to even better gradient estimation~\cite{oosterhuis2016probabilistic}.

In contrast to Offline \acs{LTR} \cite{Joachims2002, burges2010ranknet}, work in \ac{OLTR} has only considered optimizing linear combinations of ranking features~\cite{yue09:inter, hofmann12:balancing}.
Recent research has focused on improving the gradient estimation of the \ac{MGD} algorithm \cite{schuth2016mgd, oosterhuis2016probabilistic}.\
We argue this focus is a consequence of the speed-quality tradeoff; since \acs{OLTR} algorithms are evaluated by the final model they produce (i.e., offline performance) and the user experience during training (i.e., online performance), improvements should not sacrifice either of these aspects. Unfortunately, every model falls on one side of the tradeoff. For instance, more complex models like regression forests or neural networks \cite{burges2010ranknet} are very prominent in offline \acs{LTR} but they require much larger amounts of training data than for instance a simpler linear model. Thus initially more users will be shown inferior rankings when training such a complex model . Although such models may eventually find the optimal rankings, they sacrifice the user experience during training and thus will not beat the \acs{MGD} baseline in online performance. Our solution to this tradeoff is meant to stimulate the exploration of a wider range of ranking models in \acs{OLTR}.

\subsection{Multileave gradient descent}
We build on the \acl{MGD} algorithm \cite{schuth2016mgd}; see Algorithm~\ref{alg:mgd}. Briefly, at all times the algorithm has a \emph{current best} ranker $\mathbf{w}^t_0$ that is the estimate of the optimal ranker at timestep $t$. Initially, this model starts at the root $\mathbf{w}^0_0=0$, then after each issued query, another $n$ rankers $\mathbf{w}_t^n$ are sampled from the unit sphere around the \emph{current best} ranker (Line~\ref{line:mgd:candidate}). These sampled rankers are candidates: slight variations of the \emph{current best}; \acs{MGD} tries to infer if these variations are an improvement and updates accordingly. The candidates produce rankings for the query, which are combined into a single multileaved result list, e.g., by using Probabilistic Multileaving \cite{oosterhuis2016probabilistic, schuth2015probabilistic} (Line~\ref{line:mgd:multileave}). The resulting result list is displayed to the user and clicks are observed (Line \ref{line:mgd:click}); from the clicks the rankers preferred over the \emph{current best} are inferred (Line~\ref{line:mgd:infer}).
If none of the other rankers is preferred the \emph{current best} is kept, otherwise the model takes a $\eta$ step towards the mean of the winning rankers (Line~\ref{line:mgd:update}). After the model has been updated, the algorithm waits for the next query to repeat the process.

\begin{algorithm}[t]
\caption{Multileave Gradient Descent (MGD)~\cite{schuth2016mgd}.} 
\label{alg:mgd}
\begin{algorithmic}[1]
\STATE \textbf{Input}: $n$, $\delta$, $\mathbf{w}^0_0$, $\eta$
\FOR{$t \leftarrow 1..\infty$ }
	\STATE $q_t \leftarrow \mathit{receive\_query}(t)$\hfill \textit{\small // obtain a query from a user} \label{line:mgd:query}
	\STATE $\mathbf{l}_0 \leftarrow \mathit{generate\_list}(\mathbf{w}^0_t,q_t)$ \hfill \textit{\small // ranking of current best} 
	\FOR{$i \leftarrow 1..n$}\label{line:mgd:loopstart}
		\STATE $\mathbf{u}^i_t \leftarrow \mathit{sample\_unit\_vector}()$ \label{line:mgd:unitsphere}
		\STATE $\mathbf{w}_t^i \leftarrow  \mathbf{w}^0_t + \delta \mathbf{u}^i_t  $  \hfill \textit{\small // create a candidate ranker} \label{line:mgd:candidate}
		\STATE $\mathbf{l}_i \leftarrow \mathit{generate}\_list(\mathbf{w_t^i},q_t)$ \hfill \textit{\small // exploratory ranking} \label{line:mgd:loopstop}
	\ENDFOR
	\STATE $\mathbf{m}_t \leftarrow \mathit{multileave}(\mathbf{l})$\hfill \textit{\small // multileaving} \label{line:mgd:multileave}
	\STATE $\mathbf{c}_t \leftarrow \mathit{receive\_clicks}(\mathbf{m}_t)$\hfill \textit{\small // show multileaving to the user} \label{line:mgd:click}
	\STATE $\mathbf{b}_t \leftarrow \mathit{infer\_preferences}(\mathbf{l},\mathbf{m}_t,\mathbf{c}_t)$ \hfill \textit{\small // winning rankers} \label{line:mgd:infer}
	\STATE $\mathbf{w}^0_{t+1} \leftarrow \mathbf{w}^0_{t} + \eta \frac{1}{|\mathbf{b}_t|}\sum_{j \in \mathbf{b}_t} \mathbf{u}^j_t$ \label{line:mgd:update} \hfill \textit{\small // winning set may be empty}
\ENDFOR
\end{algorithmic}
\end{algorithm}

\subsection{Combining models in machine learning}

Combining models is a prevalent approach in machine learning~\cite{bishop2006pattern}; often, this is done by averaging the predictions of a set of models~\cite{breiman2001random}. Alternatively, some methods select which model to use based on the input variables~\cite{jacobs1991adaptive}. A set of multiple models whose output is averaged is called a committee, a concept that can be applied in different ways. The simplest way is by \emph{bagging}: training different models on bootstrapped datasets and taking the mean of their predictions \cite{breiman1996bagging}.
A more powerful committee technique is boosting \cite{freund1996experiments}, which trains models in sequence. Each model is trained on a weighted form of the dataset where the weights of a datapoint depend on the performance of the committee thus far. Hence, training will give more weight to points that are misclassified by the previous models. When the committee is complete their predictions are combined using a weighted voting scheme. This form of boosting is applicable to supervised classification \cite{freund1996experiments} and regression \cite{friedman2000additive}; it has also been used extensively in offline \acs{LTR}, e.g., in LambdaMART \cite{burges2010ranknet}. The main difference with our approach and ensemble methods is that their aim is to reduce the final error of the committee. None of the ensemble methods are based around user interactions; hence, none deal with the speed-quality tradeoff.


\smallskip\noindent%
On top of the related work discussed above we contribute the following: a novel \ac{OLTR} method that ranks based on feature similarities with example documents. This is the first \ac{OLTR} model that is not a direct linear model. Furthermore, we introduce a novel \ac{OLTR} algorithm that combines multiple ranking models, unlike the model combining methods discussed before this method does not combine the output of two models. Instead, different parts of the learning process are assigned to the models that are expected to perform best during that period, i.e., a model that requires less data will perform better in the initial phase of learning. This makes it the first algorithm that uses multiple ranking models to increase the user experience during learning.


\section{Sim-MGD: A Fast \ac{OLTR} Model Based on Document Feature Similarity}
\label{sec:simmgd}
In this section we introduce a novel ranking model for \ac{OLTR}, by basing result lists on feature similarities with reference documents it learns more rapidly than \ac{MGD}. However, as predicted by the speed-quality tradeoff, the increase in speed sacrifices some of the expressiveness of the model; Section~\ref{sec:cmgd} provides a method for dealing with this tradeoff.

Previous work in \ac{OLTR} has only considered optimizing linear combinations of features of documents.\footnote{Though learning the parameters for individual ranking features was researched \cite{schuth2014optimizing}.} Let $\mathbf{w}$ be the set of weights that is learned and $\mathbf{d}$ the feature representation of a query document  pair. Then a document is ranked according to the score of:
\begin{align}
R_\mathit{MGD}(\mathbf{d}) = \sum_i w_i d_i. \label{eq:mgdmodel}
\end{align}
There are several properties of the \acs{LTR} problem that this model does not make use of. For instance, almost all~\ac{LTR} features are relevance signals (e.g., BM25 or PageRank), so it is very unlikely that any should be weighted negatively. However, \ac{MGD} does not consider this when exploring; it may even consider a completely negative ranker.

As an alternative, we propose a ranking model based on the assumption that relevant documents have similar features.
Here, a set of document-query pairs $\mathbf{D}_M = \{\mathbf{d}_1,\ldots,\mathbf{d}_m\}$ is used as reference points, documents are then ranked based on their weighted similarity to those in the set:
\begin{align}
R_\mathit{sim}(\mathbf{d}) = \sum^M_{m=1} \frac{w_m}{|\mathbf{d}_m|} \mathbf{d}^T\mathbf{d}_m \label{eq:simmodel}
\end{align}
where the documents in $\mathbf{D}_M$ are $L_2$-normalized.
Since this model consists of a linear combination, optimizing its weights $\mathbf{w}$ is straightforward with the existing \ac{MGD} (Algorithm~\ref{alg:mgd}) or with our novel algorithm \ac{C-MGD} (Algorithm~\ref{alg:cmgd}) to be introduced below. For clarity we have displayed \ac{MGD} optimizing the similarity model \ac{Sim-MGD} in Algorithm~\ref{alg:simmgd}. Unlike \ac{MGD}, \ac{Sim-MGD} requires a collection of document-query pairs from which the set $\mathbf{D}_m$ is sampled (Line~\ref{line:simmgd:sample}). \ac{Sim-MGD} is still initialized with $\mathbf{w}^0_0 = 0$ but the number of weights is now determined by the size of the reference set $M$. For each query that is received, a result list is created by the \emph{current best} ranker (Line~\ref{line:simmgd:cbranking}); here, the ranker is defined by the weights $\mathbf{w}^0_t$ and the set $\mathbf{D}_M$ according to Equation~\ref{eq:simmodel}. Then $n$ candidates are sampled around the \emph{current best} ranker (Line~\ref{line:simmgd:candidate}) and their result lists are also created using Equation~\ref{eq:simmodel} (Line~\ref{line:simmgd:loopstop}). The result lists are combined into a multileaving and presented to the user (Line~\ref{line:simmgd:multileave}--\ref{line:simmgd:click}); if preferences are inferred from their interactions with the displayed result list, the \emph{current best} ranker is updated accordingly (Line~\ref{line:simmgd:infer}--\ref{line:simmgd:update}).

The intuition behind \ac{Sim-MGD} is that it is easier to base a result list on good or bad examples than it is to discover how each feature should be weighed. Moreover, \ac{MGD} optimizes faster in spaces with a lower dimensionality \cite{yue09:inter}; thus, a small number of reference documents $M$ speeds up learning further.
In spite of this speedup, the similarity model is less expressive than the standard linear model (Equation~\ref{eq:mgdmodel}). Regardless of $\mathbf{D}_M$, the similarity model can always be rewritten to a linear model:
\begin{align}
R(\mathbf{d}) = \sum^M_{m=1} \frac{w_m}{|\mathbf{d}_m|} \mathbf{d}^T\mathbf{d}_m
=  \mathbf{d}^T \sum^M_{m=1} \frac{w_m}{|\mathbf{d}_m|} \mathbf{d}_m. \label{eq:simtolin}
\end{align}
However, not every linear model can necessarily be rewritten as a similarity model, especially if the reference set $\mathbf{D}_M$ is small. Thus the space of models is limited by $\mathbf{D}_M$, providing faster learning but potentially excluding the optimal ranker. Therefore, the similarity model falls on the speed side of the speed-quality tradeoff.

For this paper, different sampling methods for creating $\mathbf{D}_M$ (Line~\ref{line:simmgd:sample}) are investigated. First, a uniform sampling, expected to cover all documents evenly, is considered. Additionally, k-means clustering is used, where $k=M$ and the centroid of each cluster is used as a reference document; this increases the chance of representing all different document types in the reference set. 

\ac{Sim-MGD} is expected to learn faster and provide a better initial user experience than \ac{MGD}. However, it is less expressive
and is thus expected to converge at an inferior optimum. Again, without the use of \ac{C-MGD} the similarity model falls on the speed side of the speed-quality tradeoff.

\begin{algorithm}[t]
\caption{\ac{MGD} with the Similarity Model (\ac{Sim-MGD}).} 
\label{alg:simmgd}
\begin{algorithmic}[1]
\STATE \textbf{Input}: $\mathbf{C}$, $M$, $n$, $\delta$, $\mathbf{w}^0_0$, $\eta$
\STATE $\mathbf{D}_M \leftarrow \{\mathbf{d}_0,\ldots,\mathbf{d}_m\} \sim \textit{sample}(\mathbf{C})$  \hfill \textit{\small// sample reference documents}  \label{line:simmgd:sample}
\FOR{$t \leftarrow 1..\infty$ }
	\STATE $q_t \leftarrow \mathit{receive\_query}(t)$\hfill \textit{\small // obtain a query from a user} \label{line:simmgd:query}
	\STATE $\mathbf{l}_0 \leftarrow \mathit{generate\_list}(\mathbf{w}^0_t,q_t,\mathbf{D}_M)$ \hfill \textit{\small// exploitive ranking (Eq.~\ref{eq:simmodel})}  \label{line:simmgd:cbranking}
	\FOR{$i \leftarrow 1..n$}\label{line:simmgd:loopstart}
		\STATE $\mathbf{u}^i_t \leftarrow \mathit{sample\_unit\_vector}()$ \label{line:simmgd:unitsphere}
		\STATE $\mathbf{w}_t^i \leftarrow  \mathbf{w}^0_t + \delta \mathbf{u}^i_t  $  \hfill \textit{\small // create a candidate ranker} \label{line:simmgd:candidate}
		\STATE $\mathbf{l}_i \leftarrow \mathit{generate}\_list(\mathbf{w_t^i},q_t,\mathbf{D}_M )$ \hfill \textit{\small// exploratory ranking (Eq.~\ref{eq:simmodel})} \label{line:simmgd:loopstop}
	\ENDFOR
	\STATE $\mathbf{m}_t \leftarrow \mathit{multileave}(\mathbf{l})$\hfill \textit{\small // multileaving} \label{line:simmgd:multileave}
	\STATE $\mathbf{c}_t \leftarrow \mathit{receive\_clicks}(\mathbf{m}_t)$\hfill \textit{\small // show multileaving to the user} \label{line:simmgd:click}
	\STATE $\mathbf{b}_t \leftarrow \mathit{infer\_preferences}(\mathbf{l},\mathbf{m}_t,\mathbf{c}_t)$ \hfill \textit{\small // winning rankers} \label{line:simmgd:infer}
	\STATE $\mathbf{w}^0_{t+1} \leftarrow \mathbf{w}^0_{t} + \eta \frac{1}{|\mathbf{b}_t|}\sum_{j \in \mathbf{b}_t} \mathbf{u}^j_t$ \label{line:simmgd:update} \hfill \textit{\small // winning set may be empty}
\ENDFOR
\end{algorithmic}
\end{algorithm}

%

\section{C-MGD: Combining \ac{OLTR} Models as a Cascade}
\label{sec:cmgd}

We aim to combine the initial learning speed of one model and the final convergence of another. This provides the best performance and user experience in the short and long term. Our proposed algorithm makes use of a cascade: initially it optimizes the faster model by letting it interact with the users until convergence is detected. At this point, the learning speed of the faster model will no longer be of advantage as the model is oscillating around a (local) optimum. Furthermore, it is very likely that a better optimum exists in a more expressive model space, especially if the faster model is relatively simple.
To make use of this likelihood, optimization is continued using a more complex model that is initialized with the first model. If this switch is made appropriately, the advantages of both models are combined: a fast initial learning speed and convergence at a better optimum. We call this algorithm \acfi{C-MGD}; before it is detailed, we discuss the main challenges of switching between models during learning.

\subsection{Detecting convergence}
\ac{C-MGD} has to detect convergence during optimization. After sufficiently many interactions, the performance of \ac{MGD} plateaus \cite{oosterhuis2016probabilistic, schuth2016mgd}. However, in the online setting there is no validation set to verify this. Instead, convergence of the model itself can be measured by looking at how much it has changed over a recent period of time. If the ranker has barely changed,
then either the estimated gradient is oscillating around a point in the model space, or few of the clicks prefer the candidates that \ac{MGD} has proposed. Both cases are indicative of finding a (local) optimum. Correspondingly, during \ac{MGD} optimization the convergence of a model $\mathbf{w}^t$ at timestep $t$ can be assumed if it has not changed substantially during the past $h$ iterations. \ac{C-MGD} considers a change significant if the cosine similarity between the current model and the model of $h$ iterations earlier exceeds a chosen threshold $\epsilon$:
\begin{align}
\frac{\mathbf{w}^{t} \cdot \mathbf{w}^{t-h}}{\|\mathbf{w}^{t}\|\cdot\|\mathbf{w}^{t-h}\|} < 1 - \epsilon.
\end{align}
The cosine similarity is appropriate here since linear combinations are unique by their direction and not their norm. Since scaling the weights of a model produces the same rankings, i.e., for a document pair $\{ \mathbf{d}_i , \mathbf{d}_j\}$:
\begin{align}
\mathbf{w} \cdot \mathbf{d}_i > \mathbf{w}\cdot \mathbf{d}_j \rightarrow \beta \mathbf{w}\cdot \mathbf{d}_i > \beta \mathbf{w}\cdot \mathbf{d}_j.
\end{align}
Therefore, a minor change in the cosine similarity indicates that the model creates rankings that are only slightly different.

\subsection{Difference in confidence}
\ac{C-MGD} has to account for the difference in confidence when changing model space. 
Convergence in the simpler model space gives \ac{C-MGD} confidence that an optimum was found, but some of this confidence is lost when switching model spaces since a lot of the new space has not been explored.
\ac{MGD}'s confidence is indicated by the norm of its model's weights, which increases if a preference in the same direction is repeatedly found.
Consequently, when initializing the subsequent model \ac{C-MGD} has to renormalize for the difference in confidence due to switching model spaces.
This is not trivial as it affects exploration, since the norm determines how dissimilar the sampled candidates will be.
If the norm is set too low it will continue by exploring a large region of model space, thus neglecting the learning done by the previous model. But if \ac{C-MGD} starts with a norm that is too large it will continue with so little exploration that it may not find the new optimum in a reasonable amount of time.

Directly measuring confidence is not possible in the online setting. Instead, rescaling is estimated from the difference in dimensionality of the models:
\begin{align}
\|\mathbf{w}_\textit{complex}\| = \|\mathbf{w}_\textit{simple}\| \cdot \frac{\sqrt{D_\textit{simple}}}{\sqrt{D_\textit{complex}}},
\end{align}
where $D_\textit{simple}$ and $D_\textit{complex}$ are the dimensionality of the simple and complex model respectively.
In line with the regret bounds found by \citet{yue09:inter}, the algorithm's confidence decreases when more parameters are introduced.

\if
\else
Lastly, another issue with swapping models is that some complex model cannot be initialized with the simpler model. For instance, if a non-linear kernel is used in the similarity model then there is no equivalent linear model. This issue can be solved by interpreting as the simpler model as another ranking signal. For the similarity model a new feature $\phi_0(\mathbf{d}) = R_{sim}(\mathbf{d})$ can be added to a linear model with all zero weights except for the corresponding weight $w_0$. Since most models consists of linear combinations this approach is usually applicable, i.e. regression forests and neural networks both are linear combinations of either trees or nodes, adding the previous model to this combination is straightforward. As a result any combination of simple and complex models can be used by ultimately optimizing their linear combination.
\fi

\subsection{A walkthrough of \ac{C-MGD}}
Finally, \ac{C-MGD} is formulated in Algorithm~\ref{alg:cmgd}. As input, \ac{C-MGD} takes two ranking models $R_\mathit{simple}$ and $R_\mathit{complex}$ with dimensionalities $D_\mathit{simple}$ and $D_\mathit{complex}$.
\ac{C-MGD} will optimize its \emph{current best} weights $\mathbf{w}^0_t$ for its current model $R_*$. Initially, $R_*$ is set to the fast learner: $R_\mathit{simple}$ (Line~\ref{line:cmgd:initmodel}). Then, for each incoming query (Line~\ref{line:cmgd:query}) the ranking of the current model ($R_{*},\mathbf{w}^0_t$) is generated (Line~\ref{line:cmgd:generatelist}). Subsequently, $n$ candidates are sampled from the unit sphere around the current weights and the ranking of each candidate is generated (Line~\ref{line:cmgd:sample}--\ref{line:cmgd:candlist}). All of the rankings are then combined into a single multileaving \cite{schuth2015probabilistic} and displayed to the user (Line~\ref{line:cmgd:multileave}--\ref{line:cmgd:clicks}). Based on the clicks of the user, a preference between the candidates and the \emph{current best} can be inferred (Line~\ref{line:cmgd:pref}). If some candidates are preferred over the \emph{current best}, an update is performed to take an $\eta$ step towards them (Line~\ref{line:cmgd:update}). Otherwise, the \emph{current best} weights will be carried over to the next iteration. At this point \ac{C-MGD} will check for convergence by comparing the cosine similarity between the \emph{current best} and the weights from $h$ iterations before: $w^0_{t-h}$ (Line~\ref{line:cmgd:convergence}).
If convergence is detected, \ac{C-MGD} switches to the complex model (Line~\ref{line:cmgd:swap}) and the \emph{current best} weights are converted to the new model space (Line~\ref{line:cmgd:project}).
The weights now have to be renormalized to account for the change in model space and rescaled for the difference in confidence (Line~\ref{line:cmgd:rescaling}). 
Optimization now continues without the check for convergence. 

The result is an algorithm that optimizes a cascade of two models, combining the advantages of both. For this study we only considered a cascade of two models, extending this approach to a larger number is straightforward.

\begin{algorithm}[t]
\caption{Cascading Multileave Gradient Descent (C-MGD).} 
\label{alg:cmgd}
\begin{algorithmic}[1]
\STATE \textbf{Input}: $n$, $\delta$, $\mathbf{w}^0_0$, $h$, $\epsilon$, $R_\mathit{simple}$,  $R_\mathit{complex}$, $D_\mathit{simple}$,  $D_\mathit{complex}$
\STATE $R_{*} \gets R_\textit{simple}$ \label{line:cmgd:initmodel}
\FOR{$t \leftarrow 1..\infty$ }
	\STATE $q_t \leftarrow \mathit{receive\_query}(t)$\hfill \textit{\small // obtain a query from a user} \label{line:cmgd:query}
	\STATE $\mathbf{l}_0 \leftarrow \mathit{generate\_list}(R_{*},\mathbf{w}^0_t,q_t)$ \hfill \textit{\small // ranking of current best} \label{line:cmgd:generatelist}
	\FOR{$i \leftarrow 1...n$}
		\STATE $\mathbf{u}^i_t \leftarrow \mathit{sample\_unit\_vector}()$ \label{line:cmgd:sample}
		\STATE $\mathbf{w}_t^i \leftarrow  \mathbf{w}^0_t + \delta \mathbf{u}^i_t  $  \hfill \textit{\small // create a candidate ranker}
		\STATE $\mathbf{l}_i \leftarrow generate\_list(\mathbf{w_t^i},q_t,R_{*})$ \hfill \textit{\small // exploratory ranking} \label{line:cmgd:candlist}
	\ENDFOR
	\STATE $\mathbf{m}_t, \mathbf{t}_t \leftarrow \mathit{multileave}(\mathbf{l})$\hfill \textit{\small // multileaving and teams} \label{line:cmgd:multileave}
	\STATE $\mathbf{c}_t \leftarrow \mathit{receive\_clicks}(\mathbf{m}_t)$\hfill \textit{\small // show multileaving to the user} \label{line:cmgd:clicks}
	\STATE $\mathbf{b}_t \leftarrow \mathit{infer\_preferences}(\mathbf{t}_t,\mathbf{c}_t)$ \hfill \textit{\small // winning candidates} \label{line:cmgd:pref}
	\STATE $\mathbf{w}^0_{t+1} \leftarrow \mathbf{w}^0_{t} + \eta \frac{1}{|\mathbf{b}_t|}\sum_{j \in \mathbf{b}_t} \mathbf{u}^j_t$  \hfill \textit{\small // winning set may be empty} \label{line:cmgd:update}
	\IF{$t \geq h \land R_{*} = R_\mathit{simple} \land 1 - \cos(\mathbf{w}^0_{t+1},\mathbf{w}^0_{t-h}) < \epsilon$} \label{line:cmgd:convergence}
		\STATE $R_{*} \gets R_\mathit{complex}$ \label{line:cmgd:swap}
		\STATE $\mathbf{w}' \gets \mathit{convert}_{R_\mathit{simple} \rightarrow R_\mathit{complex}}(\mathbf{w}_{t+1})$ \hfill \textit{\small // new model space} \label{line:cmgd:project}
		\STATE $\mathbf{w}_{t+1} \gets \mathbf{w}' \cdot \frac{\|\mathbf{w}_{t+1}\|}{\|\mathbf{w}'\|} \cdot \frac{\sqrt{D_\mathit{simple}}}{\sqrt{D_\mathit{complex}}}$\label{line:cmgd:rescaling}
	\ENDIF
\ENDFOR
\end{algorithmic}
\end{algorithm}


\section{Experiments}
\label{sec:experiments}

This section describes the experiments we run to answer the research questions posed in Section~\ref{sec:intro}.
Firstly (\ref{rq:simmgd}), we are interested in whether \ac{Sim-MGD} provides a better user experience, i.e., online performance, than \ac{MGD}. Secondly (\ref{rq:cmgd}), we wish to know if \ac{C-MGD} is capable of dealing with the speed-quality tradeoff, that is, whether \ac{C-MGD} can provide the improved user experience of \ac{Sim-MGD} (online performance) while also having the optimal convergence of \ac{MGD} (offline performance).

Every experiment below is based around a stream of independent queries coming from users. The system responds to a query by presenting a list of documents to the user in an impression. The user may or may not interact with the list by clicking on one or more documents. The queries and documents come from static datasets (Section~\ref{sec:experiments:datasets}), users are simulated using click models (Section~\ref{sec:experiments:users}).
Our experiments are described in Section~\ref{sec:experiments:runs} 
and our metrics in Section~\ref{sec:experiments:evaluation}.

\subsection{Datasets}
\label{sec:experiments:datasets}

Our experiments are performed over eleven publicly available \acs{OLTR} datasets with varying sizes and representing different search tasks. Each dataset consists of a set of queries and a set of corresponding documents for every query. While queries are represented only by their identifiers, feature representations and relevance labels are available for every document-query pair. Relevance labels are graded differently by the datasets depending on the task they model; for instance, the navigational datasets have binary labels for not relevant (0) and relevant (1), whereas most informational tasks have labels ranging from not relevant (0) to perfect relevancy (5).
Every dataset is divided in training, validation and test partitions.

The first publicly available \emph{Learning to Rank} datasets are distributed as LETOR 3.0 and 4.0~\cite{letor}; they use representations of 45, 46, or 64 features, respectively, that encode ranking models such as TF.IDF, BM25, Language Modelling, PageRank, and HITS on different parts of the documents. The datasets in LETOR are divided by their tasks, most of which come from the TREC Web Tracks between 2003 and 2008 \cite{craswell2003overview,clarke2009overview}: \emph{HP\-2003}, \emph{HP\-2004}, \emph{NP\-2003}, and \emph{NP\-2004} are based on navigational tasks
; both \emph{TD\-2003} and \emph{TD\-2004} implement the informational task of topic distillation. \emph{HP2003, HP2004, NP2003, NP2004, TD2003} and \emph{TD2004} each contain between 50 and 150 queries and 1,000 judged documents per query. The \emph{OH\-SU\-MED} dataset is based on a query log of the search engine on the MedLine abstract database, and contains 106 queries. Lastly, the two most recent datasets \emph{MQ2007} and \emph{MQ2008} were based on the Million Query Track \cite{allan2007million} and consist of 1700 and 800 queries, respectively, but have far fewer assessed documents per query. 

In 2010 Microsoft released the \emph{MSLR-WEB30k} and \emph{MLSR-WEB10K} \cite{Qin2013Letor}, the former consists of 30,000 queries obtained from a retired labelling set of a commercial web search engine (Bing), the latter is a subsampling of 10,000 queries from the former dataset. The datasets uses 136 features to represent its documents, each query has around 125 assessed documents. 
 For practical reasons only \emph{MLSR-WEB10K} was used for this paper.

Lastly, also in 2010 Yahoo! organised a public Learning to Rank Challenge \cite{Chapelle2011} with an accompanying dataset. This set consist of 709,877 documents encoded in 700 features and sampled from query logs of the Yahoo! search engine spanning  29,921 queries.

\subsection{Simulating user behavior}
\label{sec:experiments:users}

\begin{table}[tb]
\caption{Instantiations of Cascading Click Models~\cite{guo09:efficient} as used for simulating user behavior in experiments.}
\centering
\begin{tabularx}{\columnwidth}{ l c c c c c c c c c c }
\toprule

& \multicolumn{5}{c}{\small $P(\mathit{click}=1\mid R)$} & \multicolumn{5}{c}{\small $P(\mathit{stop}=1\mid R)$} \\
\small $R$ & \small \emph{$ 0$} & \small \emph{$ 1$}  & \small \emph{$ 2$} & \small \emph{$ 3$} & \small \emph{$ 4$}
 & \small \emph{$0$} & \small \emph{$ 1$} & \small \emph{$ 2$} & \small \emph{$ 3$} & \small \emph{$ 4$} \\
\midrule
\small \emph{perf} & \small 0.0 & \small 0.2 & \small 0.4 & \small 0.8 & \small 1.0 & \small 0.0 & \small 0.0 & \small 0.0 & \small 0.0 & \small 0.0 \\
\small \emph{nav} & \small ~~0.05 & \small 0.3 & \small 0.5 & \small 0.7 & \small ~~0.95 & \small 0.2 & \small 0.3 & \small 0.5 & \small 0.7 & \small 0.9 \\
\small \emph{inf} & \small 0.4 & \small 0.6 & \small 0.7 & \small 0.8 & \small 0.9 & \small 0.1 & \small 0.2 & \small 0.3 & \small 0.4 & \small 0.5 \\
\bottomrule
\end{tabularx}
\label{tab:clickmodels}
\end{table}

Users are simulated using the standard setup for \acs{OLTR} simulations \cite{Hofmann2013a,schuth2016mgd,oosterhuis2016probabilistic}.  First, a user issues a query  simulated by uniformly sampling a query from the static dataset. Subsequently, the algorithm decides the result list of documents to display. The behavior of the user after it receives this result list is simulated using a \emph{cascade click model}~\cite{chuklin-click-2015,guo09:efficient}. This model assumes a user to examine the documents of the result list in their displayed order. For each document that is considered the user decides whether it warrants a click. This is modelled as the conditional probability $P(click=1\mid R)$, where $R$ is the relevance label provided by the dataset. Accordingly, \emph{cascade click model} instantiations increase the probability of a click with the degree of the relevance label. After the user has clicked on a document, their information need may be satisfied, otherwise they will continue by considering the remaining documents. The probability of the user not examining more documents after clicking is modeled as $P(stop=1\mid R)$, 
where it is more likely that the user is satisfied from a very relevant document. For this paper $\kappa=10$ documents are displayed to the user at each impression.

Table~\ref{tab:clickmodels} lists the three instantiations of cascade click models that were used for this paper. The first models a \emph{perfect} user that considers every document and clicks on all relevant documents and nothing else.
Secondly, the \emph{navigational} instantiation models a user performing a navigational task and is mostly looking for a single highly relevant document. Finally, the \emph{informational} instantiation models a user without a very specific information need that typically clicks on multiple documents. 
These three models have increasing levels of noise, as the behavior of each depends less on the relevance labels of the displayed documents.

\subsection{Experimental runs}
\label{sec:experiments:runs}

As a baseline, Probabilistic-\acs{MGD}~\cite{oosterhuis2016probabilistic} is used. Based on previous work this study uses $n=19$ candidates per iteration sampled from the unit sphere with $\delta=1$; updates are performed with $\eta = 0.01$ and weights are intialized as $\mathbf{w}^0_0 = \mathbf{0}$ \cite{yue09:inter, hofmann11:balancing, schuth2016mgd, oosterhuis2016probabilistic}. All runs are run over 10,000 impressions.
Probabilistic Multileaving inferences are computed using a sample-based method \cite{schuth2015probabilistic}, where the number of document assignments sampled for every inference is 10,000 \cite{oosterhuis2016probabilistic}.

\ac{Sim-MGD} uses $M=50$ reference documents that are selected from the training set at the start of each run. 
The choice for $M=50$ was based on preliminary results on the evaluation sets.
Two selection methods are investigated: uniform sampling and k-means clustering. The clustering method uses $k=M$, i.e., producing a reference document for every cluster it finds. The expectation is that \ac{Sim-MGD} has a higher learning speed but is less expressive than \ac{MGD}, thus, we expect to see a substantial increase in online performance but a decrease in offline performance compared to \ac{MGD}.
Clustering is expected to provide reference documents that cover all kinds of documents better, potentially resulting in a further increase of online performance and a lower standard deviation compared to uniform sampling.

Finally, to evaluate whether \ac{C-MGD} can successfully combine speed and quality of two models, \ac{C-MGD} is run with \ac{Sim-MGD} as $R_\textit{simple}$ (Equation~\ref{eq:simmodel}) and the linear model as $R_\textit{complex}$ (Equation~\ref{eq:mgdmodel}). If the cascade can successfully swap models then we expect to see no significant decrease in offline performance but a substantial increase in online performance compared to \ac{MGD}. When comparing to \ac{Sim-MGD} we expect a significant increase in offline performance due to \ac{C-MGD}'s ability to switch models. However, it is very likely that a slight decrease in online performance is observed, since the change of model space introduces more exploration. Lastly, the reference document selection methods are expected to have the same effects on \ac{C-MGD} as they have on \ac{Sim-MGD}.

\subsection{Metrics and tests}
\label{sec:experiments:evaluation}

The task in \acs{OLTR} consists of two parts: a ranker has to be optimized and users have to be attended during optimization. Accordingly, both aspects are evaluated separately. 

\emph{Offline performance} considers the quality of the learned model by taking the average NDCG score of the \emph{current best ranker} over a held-out set.
Performance is assessed using the NDCG~\citep{jarvelin2002:cumulated} metric:
\begin{align}
\mathit{NDCG} = \sum^{\kappa}_{i=1} \frac{2^{\mathit{rel}(\mathbf{r}[i])}-1}{\log_2(i+1)} \mathit{iDCG}^{-1}.
\end{align}
This metric calculates the Discounted Cumulative Gain (DCG) over the relevance labels $\mathit{rel}(\mathbf{r}[i])$ for each document in the top $\kappa$ of a ranking. Subsequently, this is normalized by the maximal DCG possible for a query: the ideal DCG (iDCG). 
This results in Normalized DCG (NDCG) which measures the quality of a single ranked list of documents. 
Offline performance is averaged over a held-out set after 10,000 impressions to give an indication at what performance the algorithms converge.

Conversely, the user experience during training is essential as well, since deterring users during training would compromise the purpose of the system. \emph{Online performance} is assessed by computing the cumulative NDCG of the rankings shown to the users \cite{Hofmann2013a, sutton1998:introduction}. For $T$ successive queries this is the discounted sum:
\begin{align}
\mathit{Online\_Performance} = \sum_{t=1}^T \mathit{NDCG}(\mathbf{m}_t) \cdot \gamma^{(t-1)}
\end{align}
where $\mathbf{m}_t$ is the ranking displayed to the user at timestep $t$. This metric is common in \emph{online learning} and can be interpreted as the expected reward with $\gamma$ as the probability that another query will be issued.
For \emph{online performance} a discount factor of $\gamma = 0.9995$ was chosen so that queries beyond the horizon of 10,000 queries have a less than $1\%$ impact \cite{oosterhuis2016probabilistic}.

Finally, all  runs are repeated 125 times, spread evenly over the dataset's folds; results for each run are averaged and a two tailed Student's t-test is used to verify whether differences are statistically significant \cite{zimmerman1987comparative}. In total, our experiments are based on over 200 million user impressions.


\section{Results and Analysis}
\label{sec:results}

\begin{table*}[tb]
\centering
\caption{Online performance (Discounted Cumulative NDCG, Section~\ref{sec:experiments:evaluation}) for different instantiations of CCM (Table~\ref{tab:clickmodels}). The standard deviation is shown in brackets, bold values indicate the highest performance per dataset and click model, significant improvements and losses over the \acs{MGD} baseline are indicated by  \enkelop\ (p $<$ 0.05) and \dubbelop\ (p $<$ 0.01) and by \enkelneer\ and \dubbelneer, respectively.}
\begin{tabular*}{\textwidth}{@{\extracolsep{\fill} } l  l l l l l  }
\toprule
& \textbf{MGD}&\multicolumn{2}{c}{\textbf{\acs{Sim-MGD}}}&\multicolumn{2}{c}{\textbf{\acs{C-MGD}}}\\
\cmidrule{3-4}\cmidrule{5-6}
 & & \multicolumn{1}{l}{ \small \textbf{\small uniform}}  & \multicolumn{1}{l}{ \small \textbf{\small k-means}}  & \multicolumn{1}{l}{ \small \textbf{\small uniform}}  & \multicolumn{1}{l}{ \small \textbf{\small k-means}} \\
\midrule
& \multicolumn{5}{c}{\textit{perfect}} \\
\midrule
HP2003 & {\small 764.4 {\tiny( 16.7)}} & {\small 750.8 {\tiny( 36.2)}} {\tiny \dubbelneer} & {\small 771.1 {\tiny( 23.6)}} {\tiny \enkelop} & {\small 770.5 {\tiny( 17.9)}} {\tiny \dubbelop} & \textbf {\small 773.0 {\tiny( 15.9)}} {\tiny \dubbelop} \\
NP2003 & {\small 699.5 {\tiny( 19.5)}} & {\small 789.1 {\tiny( 17.8)}} {\tiny \dubbelop} & \textbf {\small 794.2 {\tiny( 18.7)}} {\tiny \dubbelop} & {\small 724.6 {\tiny( 16.7)}} {\tiny \dubbelop} & {\small 723.9 {\tiny( 18.6)}} {\tiny \dubbelop} \\
TD2003 & \textbf {\small 312.2 {\tiny( 20.0)}} & {\small 280.1 {\tiny( 29.4)}} {\tiny \dubbelneer} & {\small 279.1 {\tiny( 23.0)}} {\tiny \dubbelneer} & {\small 306.0 {\tiny( 21.9)}} {\tiny \enkelneer} & {\small 307.8 {\tiny( 21.7)}} \hphantom{\tiny \dubbelneer} \\
HP2004 & {\small 732.3 {\tiny( 19.0)}} & {\small 766.8 {\tiny( 28.0)}} {\tiny \dubbelop} & \textbf {\small 777.7 {\tiny( 21.6)}} {\tiny \dubbelop} & {\small 748.1 {\tiny( 20.4)}} {\tiny \dubbelop} & {\small 746.9 {\tiny( 20.1)}} {\tiny \dubbelop} \\
NP2004 & {\small 719.9 {\tiny( 17.8)}} & {\small 769.8 {\tiny( 23.9)}} {\tiny \dubbelop} & \textbf {\small 781.8 {\tiny( 20.2)}} {\tiny \dubbelop} & {\small 737.8 {\tiny( 17.6)}} {\tiny \dubbelop} & {\small 740.5 {\tiny( 15.9)}} {\tiny \dubbelop} \\
TD2004 & \textbf {\small 298.9 {\tiny( 12.5)}} & {\small 268.1 {\tiny( 19.8)}} {\tiny \dubbelneer} & {\small 267.6 {\tiny( 11.2)}} {\tiny \dubbelneer} & {\small 295.2 {\tiny( 11.3)}} {\tiny \enkelneer} & {\small 296.4 {\tiny( 11.5)}} \hphantom{\tiny \dubbelneer} \\
MQ2007 & {\small 412.5 {\tiny( 10.4)}} & \textbf {\small 448.4 {\tiny( 10.2)}} {\tiny \dubbelop} & {\small 443.1 {\tiny( 10.7)}} {\tiny \dubbelop} & {\small 423.4 {\tiny( 10.8)}} {\tiny \dubbelop} & {\small 421.1 {\tiny( 9.9)}} {\tiny \dubbelop} \\
MQ2008 & {\small 523.2 {\tiny( 15.8)}} & \textbf {\small 547.3 {\tiny( 16.2)}} {\tiny \dubbelop} & {\small 543.1 {\tiny( 16.5)}} {\tiny \dubbelop} & {\small 531.3 {\tiny( 15.1)}} {\tiny \dubbelop} & {\small 527.1 {\tiny( 15.2)}} {\tiny \enkelop} \\
MSLR-WEB10k & {\small 336.6 {\tiny( 6.3)}} & {\small 347.9 {\tiny( 6.6)}} {\tiny \dubbelop} & \textbf {\small 351.2 {\tiny( 6.5)}} {\tiny \dubbelop} & {\small 340.7 {\tiny( 6.3)}} {\tiny \dubbelop} & {\small 342.1 {\tiny( 5.9)}} {\tiny \dubbelop} \\
OHSUMED & {\small 494.8 {\tiny( 15.8)}} & {\small 483.5 {\tiny( 16.4)}} {\tiny \dubbelneer} & {\small 483.2 {\tiny( 17.3)}} {\tiny \dubbelneer} & {\small 494.4 {\tiny( 17.2)}} \hphantom{\tiny \dubbelneer} & \textbf {\small 495.4 {\tiny( 16.6)}} \hphantom{\tiny \dubbelneer} \\
Yahoo & {\small 732.1 {\tiny( 10.9)}} & {\small 773.7 {\tiny( 12.5)}} {\tiny \dubbelop} & \textbf {\small 778.6 {\tiny( 9.9)}} {\tiny \dubbelop} & {\small 741.5 {\tiny( 10.0)}} {\tiny \dubbelop} & {\small 742.7 {\tiny( 11.7)}} {\tiny \dubbelop} \\
\midrule
& \multicolumn{5}{c}{\textit{navigational}} \\
\midrule
HP2003 & {\small 701.2 {\tiny( 19.7)}} & {\small 717.3 {\tiny( 37.5)}} {\tiny \dubbelop} & \textbf {\small 734.6 {\tiny( 21.6)}} {\tiny \dubbelop} & {\small 715.6 {\tiny( 18.3)}} {\tiny \dubbelop} & {\small 718.0 {\tiny( 18.3)}} {\tiny \dubbelop} \\
NP2003 & {\small 637.6 {\tiny( 23.0)}} & {\small 765.0 {\tiny( 19.4)}} {\tiny \dubbelop} & \textbf {\small 772.4 {\tiny( 19.0)}} {\tiny \dubbelop} & {\small 684.8 {\tiny( 18.0)}} {\tiny \dubbelop} & {\small 686.8 {\tiny( 19.3)}} {\tiny \dubbelop} \\
TD2003 & \textbf {\small 272.5 {\tiny( 19.9)}} & {\small 253.9 {\tiny( 25.2)}} {\tiny \dubbelneer} & {\small 256.4 {\tiny( 22.3)}} {\tiny \dubbelneer} & {\small 265.6 {\tiny( 20.1)}} {\tiny \dubbelneer} & {\small 269.8 {\tiny( 20.8)}} \hphantom{\tiny \dubbelneer} \\
HP2004 & {\small 663.0 {\tiny( 20.9)}} & {\small 724.5 {\tiny( 30.5)}} {\tiny \dubbelop} & \textbf {\small 742.9 {\tiny( 24.4)}} {\tiny \dubbelop} & {\small 693.4 {\tiny( 24.1)}} {\tiny \dubbelop} & {\small 693.7 {\tiny( 20.7)}} {\tiny \dubbelop} \\
NP2004 & {\small 653.2 {\tiny( 20.3)}} & {\small 743.9 {\tiny( 24.3)}} {\tiny \dubbelop} & \textbf {\small 756.1 {\tiny( 21.7)}} {\tiny \dubbelop} & {\small 686.6 {\tiny( 19.5)}} {\tiny \dubbelop} & {\small 692.2 {\tiny( 17.3)}} {\tiny \dubbelop} \\
TD2004 & \textbf {\small 263.3 {\tiny( 12.6)}} & {\small 242.5 {\tiny( 16.3)}} {\tiny \dubbelneer} & {\small 243.8 {\tiny( 10.4)}} {\tiny \dubbelneer} & {\small 260.7 {\tiny( 11.9)}} \hphantom{\tiny \dubbelneer} & {\small 261.8 {\tiny( 10.4)}} \hphantom{\tiny \dubbelneer} \\
MQ2007 & {\small 385.9 {\tiny( 11.9)}} & \textbf {\small 430.5 {\tiny( 12.3)}} {\tiny \dubbelop} & {\small 424.3 {\tiny( 11.7)}} {\tiny \dubbelop} & {\small 402.4 {\tiny( 9.7)}} {\tiny \dubbelop} & {\small 402.0 {\tiny( 11.2)}} {\tiny \dubbelop} \\
MQ2008 & {\small 501.5 {\tiny( 16.3)}} & \textbf {\small 534.6 {\tiny( 14.1)}} {\tiny \dubbelop} & {\small 529.5 {\tiny( 14.4)}} {\tiny \dubbelop} & {\small 513.4 {\tiny( 15.2)}} {\tiny \dubbelop} & {\small 511.8 {\tiny( 15.3)}} {\tiny \dubbelop} \\
MSLR-WEB10k & {\small 323.2 {\tiny( 7.2)}} & {\small 335.0 {\tiny( 8.1)}} {\tiny \dubbelop} & \textbf {\small 338.2 {\tiny( 7.8)}} {\tiny \dubbelop} & {\small 327.7 {\tiny( 6.7)}} {\tiny \dubbelop} & {\small 330.0 {\tiny( 6.1)}} {\tiny \dubbelop} \\
OHSUMED & \textbf {\small 482.6 {\tiny( 15.9)}} & {\small 464.2 {\tiny( 19.4)}} {\tiny \dubbelneer} & {\small 465.0 {\tiny( 17.3)}} {\tiny \dubbelneer} & {\small 478.8 {\tiny( 16.6)}} \hphantom{\tiny \dubbelneer} & {\small 480.1 {\tiny( 15.2)}} \hphantom{\tiny \dubbelneer} \\
Yahoo & {\small 721.1 {\tiny( 14.4)}} & {\small 758.2 {\tiny( 27.1)}} {\tiny \dubbelop} & \textbf {\small 767.1 {\tiny( 22.2)}} {\tiny \dubbelop} & {\small 732.2 {\tiny( 17.3)}} {\tiny \dubbelop} & {\small 733.9 {\tiny( 18.4)}} {\tiny \dubbelop} \\
\midrule
& \multicolumn{5}{c}{\textit{informational}} \\
\midrule
HP2003 & {\small 650.9 {\tiny( 22.6)}} & {\small 680.5 {\tiny( 33.0)}} {\tiny \dubbelop} & \textbf {\small 703.7 {\tiny( 22.3)}} {\tiny \dubbelop} & {\small 673.2 {\tiny( 21.5)}} {\tiny \dubbelop} & {\small 675.6 {\tiny( 20.5)}} {\tiny \dubbelop} \\
NP2003 & {\small 603.0 {\tiny( 26.1)}} & {\small 750.8 {\tiny( 19.4)}} {\tiny \dubbelop} & \textbf {\small 757.0 {\tiny( 20.6)}} {\tiny \dubbelop} & {\small 655.7 {\tiny( 21.9)}} {\tiny \dubbelop} & {\small 657.5 {\tiny( 17.3)}} {\tiny \dubbelop} \\
TD2003 & {\small 251.6 {\tiny( 20.4)}} & {\small 247.3 {\tiny( 22.0)}} \hphantom{\tiny \dubbelneer} & {\small 248.8 {\tiny( 20.6)}} \hphantom{\tiny \dubbelneer} & \textbf {\small 257.0 {\tiny( 22.3)}} {\tiny \enkelop} & {\small 255.3 {\tiny( 21.1)}} \hphantom{\tiny \dubbelneer} \\
HP2004 & {\small 616.1 {\tiny( 25.4)}} & {\small 697.9 {\tiny( 33.9)}} {\tiny \dubbelop} & \textbf {\small 718.7 {\tiny( 23.1)}} {\tiny \dubbelop} & {\small 652.6 {\tiny( 28.9)}} {\tiny \dubbelop} & {\small 651.8 {\tiny( 23.6)}} {\tiny \dubbelop} \\
NP2004 & {\small 617.8 {\tiny( 23.4)}} & {\small 719.0 {\tiny( 26.8)}} {\tiny \dubbelop} & \textbf {\small 736.2 {\tiny( 23.7)}} {\tiny \dubbelop} & {\small 661.9 {\tiny( 21.9)}} {\tiny \dubbelop} & {\small 663.3 {\tiny( 20.1)}} {\tiny \dubbelop} \\
TD2004 & {\small 245.0 {\tiny( 15.4)}} & {\small 232.8 {\tiny( 15.9)}} {\tiny \dubbelneer} & {\small 237.0 {\tiny( 12.0)}} {\tiny \dubbelneer} & \textbf {\small 250.2 {\tiny( 13.1)}} {\tiny \dubbelop} & {\small 247.9 {\tiny( 11.5)}} \hphantom{\tiny \dubbelneer} \\
MQ2007 & {\small 377.2 {\tiny( 15.1)}} & \textbf {\small 415.4 {\tiny( 39.6)}} {\tiny \dubbelop} & {\small 404.5 {\tiny( 46.7)}} {\tiny \dubbelop} & {\small 386.5 {\tiny( 35.7)}} {\tiny \dubbelop} & {\small 387.6 {\tiny( 34.2)}} {\tiny \dubbelop} \\
MQ2008 & {\small 496.3 {\tiny( 22.0)}} & {\small 508.1 {\tiny( 60.6)}} {\tiny \enkelop} & \textbf {\small 512.9 {\tiny( 50.7)}} {\tiny \dubbelop} & {\small 489.5 {\tiny( 52.8)}} \hphantom{\tiny \dubbelneer} & {\small 491.2 {\tiny( 51.2)}} \hphantom{\tiny \dubbelneer} \\
MSLR-WEB10k & {\small 321.4 {\tiny( 8.5)}} & {\small 329.9 {\tiny( 22.9)}} {\tiny \dubbelop} & \textbf {\small 331.3 {\tiny( 25.4)}} {\tiny \dubbelop} & {\small 321.8 {\tiny( 23.9)}} \hphantom{\tiny \dubbelneer} & {\small 324.1 {\tiny( 24.2)}} \hphantom{\tiny \dubbelneer} \\
OHSUMED & {\small 474.3 {\tiny( 15.1)}} & {\small 457.5 {\tiny( 19.6)}} {\tiny \dubbelneer} & {\small 460.6 {\tiny( 18.5)}} {\tiny \dubbelneer} & {\small 473.1 {\tiny( 16.5)}} \hphantom{\tiny \dubbelneer} & \textbf {\small 474.4 {\tiny( 17.5)}} \hphantom{\tiny \dubbelneer} \\
Yahoo & {\small 707.3 {\tiny( 19.4)}} & {\small 728.8 {\tiny( 55.4)}} {\tiny \dubbelop} & \textbf {\small 734.3 {\tiny( 52.7)}} {\tiny \dubbelop} & {\small 716.9 {\tiny( 28.6)}} {\tiny \dubbelop} & {\small 714.5 {\tiny( 31.7)}} {\tiny \enkelop} \\
\bottomrule
\end{tabular*}

\label{tab:online}
\end{table*}

\begin{table*}[tb]
\centering
\caption{Offline performance (NDCG) after 10,000 impressions, notation is identical to Table~\ref{tab:online}.}
\begin{tabular*}{\textwidth}{@{\extracolsep{\fill} } l  l l l l l  }
\toprule
& \textbf{MGD}&\multicolumn{2}{c}{\textbf{\acs{Sim-MGD}}}&\multicolumn{2}{c}{\textbf{\acs{C-MGD}}}\\
\cmidrule{3-4}\cmidrule{5-6}
 & & \multicolumn{1}{l}{ \small \textbf{\small uniform}}  & \multicolumn{1}{l}{ \small \textbf{\small k-means}}  & \multicolumn{1}{l}{ \small \textbf{\small uniform}}  & \multicolumn{1}{l}{ \small \textbf{\small k-means}} \\
\midrule
& \multicolumn{5}{c}{\textit{perfect}} \\
\midrule
HP2003 & \textbf {\small 0.782 {\tiny( 0.06)}} & {\small 0.709 {\tiny( 0.06)}} {\tiny \dubbelneer} & {\small 0.720 {\tiny( 0.06)}} {\tiny \dubbelneer} & {\small 0.781 {\tiny( 0.07)}} \hphantom{\tiny \dubbelneer} & \textbf {\small 0.782 {\tiny( 0.07)}} \hphantom{\tiny \dubbelneer} \\
NP2003 & {\small 0.719 {\tiny( 0.04)}} & {\small 0.708 {\tiny( 0.04)}} {\tiny \enkelneer} & {\small 0.713 {\tiny( 0.04)}} \hphantom{\tiny \dubbelneer} & \textbf {\small 0.720 {\tiny( 0.04)}} \hphantom{\tiny \dubbelneer} & {\small 0.719 {\tiny( 0.04)}} \hphantom{\tiny \dubbelneer} \\
TD2003 & \textbf {\small 0.327 {\tiny( 0.08)}} & {\small 0.253 {\tiny( 0.09)}} {\tiny \dubbelneer} & {\small 0.243 {\tiny( 0.08)}} {\tiny \dubbelneer} & {\small 0.322 {\tiny( 0.08)}} \hphantom{\tiny \dubbelneer} & {\small 0.325 {\tiny( 0.08)}} \hphantom{\tiny \dubbelneer} \\
HP2004 & \textbf {\small 0.751 {\tiny( 0.07)}} & {\small 0.709 {\tiny( 0.09)}} {\tiny \dubbelneer} & {\small 0.714 {\tiny( 0.08)}} {\tiny \dubbelneer} & {\small 0.750 {\tiny( 0.07)}} \hphantom{\tiny \dubbelneer} & {\small 0.747 {\tiny( 0.07)}} \hphantom{\tiny \dubbelneer} \\
NP2004 & {\small 0.719 {\tiny( 0.04)}} & {\small 0.693 {\tiny( 0.07)}} {\tiny \dubbelneer} & {\small 0.698 {\tiny( 0.08)}} {\tiny \dubbelneer} & {\small 0.719 {\tiny( 0.04)}} \hphantom{\tiny \dubbelneer} & \textbf {\small 0.720 {\tiny( 0.04)}} \hphantom{\tiny \dubbelneer} \\
TD2004 & \textbf {\small 0.333 {\tiny( 0.05)}} & {\small 0.254 {\tiny( 0.03)}} {\tiny \dubbelneer} & {\small 0.254 {\tiny( 0.02)}} {\tiny \dubbelneer} & {\small 0.328 {\tiny( 0.05)}} \hphantom{\tiny \dubbelneer} & {\small 0.329 {\tiny( 0.05)}} \hphantom{\tiny \dubbelneer} \\
MQ2007 & {\small 0.406 {\tiny( 0.02)}} & {\small 0.401 {\tiny( 0.02)}} {\tiny \enkelneer} & {\small 0.395 {\tiny( 0.02)}} {\tiny \dubbelneer} & \textbf {\small 0.408 {\tiny( 0.02)}} \hphantom{\tiny \dubbelneer} & {\small 0.407 {\tiny( 0.02)}} \hphantom{\tiny \dubbelneer} \\
MQ2008 & \textbf {\small 0.493 {\tiny( 0.04)}} & {\small 0.485 {\tiny( 0.04)}} \hphantom{\tiny \dubbelneer} & {\small 0.479 {\tiny( 0.04)}} {\tiny \dubbelneer} & {\small 0.491 {\tiny( 0.04)}} \hphantom{\tiny \dubbelneer} & {\small 0.490 {\tiny( 0.04)}} \hphantom{\tiny \dubbelneer} \\
MSLR-WEB10k & \textbf {\small 0.312 {\tiny( 0.00)}} & {\small 0.301 {\tiny( 0.00)}} {\tiny \dubbelneer} & {\small 0.303 {\tiny( 0.00)}} {\tiny \dubbelneer} & \textbf {\small 0.312 {\tiny( 0.00)}} \hphantom{\tiny \dubbelneer} & \textbf {\small 0.312 {\tiny( 0.00)}} \hphantom{\tiny \dubbelneer} \\
OHSUMED & \textbf {\small 0.456 {\tiny( 0.05)}} & {\small 0.439 {\tiny( 0.05)}} {\tiny \dubbelneer} & {\small 0.442 {\tiny( 0.05)}} {\tiny \enkelneer} & {\small 0.455 {\tiny( 0.05)}} \hphantom{\tiny \dubbelneer} & {\small 0.455 {\tiny( 0.05)}} \hphantom{\tiny \dubbelneer} \\
Yahoo & \textbf {\small 0.675 {\tiny( 0.01)}} & {\small 0.656 {\tiny( 0.01)}} {\tiny \dubbelneer} & {\small 0.657 {\tiny( 0.00)}} {\tiny \dubbelneer} & {\small 0.672 {\tiny( 0.01)}} {\tiny \dubbelneer} & {\small 0.672 {\tiny( 0.01)}} {\tiny \enkelneer} \\
\midrule
& \multicolumn{5}{c}{\textit{navigational}} \\
\midrule
HP2003 & {\small 0.764 {\tiny( 0.06)}} & {\small 0.683 {\tiny( 0.07)}} {\tiny \dubbelneer} & {\small 0.690 {\tiny( 0.06)}} {\tiny \dubbelneer} & {\small 0.765 {\tiny( 0.06)}} \hphantom{\tiny \dubbelneer} & \textbf {\small 0.767 {\tiny( 0.06)}} \hphantom{\tiny \dubbelneer} \\
NP2003 & {\small 0.711 {\tiny( 0.04)}} & {\small 0.707 {\tiny( 0.04)}} \hphantom{\tiny \dubbelneer} & {\small 0.711 {\tiny( 0.04)}} \hphantom{\tiny \dubbelneer} & {\small 0.712 {\tiny( 0.04)}} \hphantom{\tiny \dubbelneer} & \textbf {\small 0.714 {\tiny( 0.04)}} \hphantom{\tiny \dubbelneer} \\
TD2003 & \textbf {\small 0.315 {\tiny( 0.09)}} & {\small 0.237 {\tiny( 0.08)}} {\tiny \dubbelneer} & {\small 0.237 {\tiny( 0.07)}} {\tiny \dubbelneer} & {\small 0.299 {\tiny( 0.09)}} \hphantom{\tiny \dubbelneer} & {\small 0.306 {\tiny( 0.10)}} \hphantom{\tiny \dubbelneer} \\
HP2004 & {\small 0.740 {\tiny( 0.07)}} & {\small 0.694 {\tiny( 0.08)}} {\tiny \dubbelneer} & {\small 0.700 {\tiny( 0.08)}} {\tiny \dubbelneer} & \textbf {\small 0.743 {\tiny( 0.07)}} \hphantom{\tiny \dubbelneer} & {\small 0.740 {\tiny( 0.07)}} \hphantom{\tiny \dubbelneer} \\
NP2004 & {\small 0.717 {\tiny( 0.04)}} & {\small 0.686 {\tiny( 0.08)}} {\tiny \dubbelneer} & {\small 0.691 {\tiny( 0.07)}} {\tiny \dubbelneer} & {\small 0.720 {\tiny( 0.05)}} \hphantom{\tiny \dubbelneer} & \textbf {\small 0.722 {\tiny( 0.05)}} \hphantom{\tiny \dubbelneer} \\
TD2004 & \textbf {\small 0.314 {\tiny( 0.05)}} & {\small 0.230 {\tiny( 0.03)}} {\tiny \dubbelneer} & {\small 0.225 {\tiny( 0.03)}} {\tiny \dubbelneer} & {\small 0.308 {\tiny( 0.04)}} \hphantom{\tiny \dubbelneer} & {\small 0.307 {\tiny( 0.04)}} \hphantom{\tiny \dubbelneer} \\
MQ2007 & {\small 0.356 {\tiny( 0.02)}} & \textbf {\small 0.387 {\tiny( 0.02)}} {\tiny \dubbelop} & {\small 0.377 {\tiny( 0.02)}} {\tiny \dubbelop} & {\small 0.364 {\tiny( 0.02)}} {\tiny \dubbelop} & {\small 0.363 {\tiny( 0.02)}} {\tiny \dubbelop} \\
MQ2008 & {\small 0.468 {\tiny( 0.03)}} & \textbf {\small 0.478 {\tiny( 0.04)}} {\tiny \enkelop} & {\small 0.470 {\tiny( 0.03)}} \hphantom{\tiny \dubbelneer} & {\small 0.470 {\tiny( 0.03)}} \hphantom{\tiny \dubbelneer} & {\small 0.467 {\tiny( 0.03)}} \hphantom{\tiny \dubbelneer} \\
MSLR-WEB10k & \textbf {\small 0.307 {\tiny( 0.00)}} & {\small 0.300 {\tiny( 0.00)}} {\tiny \dubbelneer} & {\small 0.302 {\tiny( 0.00)}} {\tiny \dubbelneer} & \textbf {\small 0.307 {\tiny( 0.00)}} \hphantom{\tiny \dubbelneer} & \textbf {\small 0.307 {\tiny( 0.00)}} \hphantom{\tiny \dubbelneer} \\
OHSUMED & \textbf {\small 0.439 {\tiny( 0.05)}} & {\small 0.405 {\tiny( 0.06)}} {\tiny \dubbelneer} & {\small 0.404 {\tiny( 0.06)}} {\tiny \dubbelneer} & {\small 0.430 {\tiny( 0.06)}} \hphantom{\tiny \dubbelneer} & {\small 0.430 {\tiny( 0.06)}} \hphantom{\tiny \dubbelneer} \\
Yahoo & \textbf {\small 0.660 {\tiny( 0.03)}} & {\small 0.637 {\tiny( 0.04)}} {\tiny \dubbelneer} & {\small 0.643 {\tiny( 0.03)}} {\tiny \dubbelneer} & {\small 0.655 {\tiny( 0.03)}} \hphantom{\tiny \dubbelneer} & {\small 0.657 {\tiny( 0.04)}} \hphantom{\tiny \dubbelneer} \\
\midrule
& \multicolumn{5}{c}{\textit{informational}} \\
\midrule
HP2003 & {\small 0.759 {\tiny( 0.06)}} & {\small 0.649 {\tiny( 0.07)}} {\tiny \dubbelneer} & {\small 0.667 {\tiny( 0.06)}} {\tiny \dubbelneer} & \textbf {\small 0.764 {\tiny( 0.07)}} \hphantom{\tiny \dubbelneer} & {\small 0.763 {\tiny( 0.07)}} \hphantom{\tiny \dubbelneer} \\
NP2003 & {\small 0.704 {\tiny( 0.05)}} & {\small 0.705 {\tiny( 0.04)}} \hphantom{\tiny \dubbelneer} & \textbf {\small 0.710 {\tiny( 0.04)}} \hphantom{\tiny \dubbelneer} & {\small 0.706 {\tiny( 0.04)}} \hphantom{\tiny \dubbelneer} & {\small 0.705 {\tiny( 0.05)}} \hphantom{\tiny \dubbelneer} \\
TD2003 & \textbf {\small 0.286 {\tiny( 0.10)}} & {\small 0.232 {\tiny( 0.08)}} {\tiny \dubbelneer} & {\small 0.241 {\tiny( 0.07)}} {\tiny \dubbelneer} & {\small 0.275 {\tiny( 0.09)}} \hphantom{\tiny \dubbelneer} & {\small 0.267 {\tiny( 0.09)}} \hphantom{\tiny \dubbelneer} \\
HP2004 & {\small 0.732 {\tiny( 0.07)}} & {\small 0.681 {\tiny( 0.07)}} {\tiny \dubbelneer} & {\small 0.688 {\tiny( 0.08)}} {\tiny \dubbelneer} & {\small 0.735 {\tiny( 0.07)}} \hphantom{\tiny \dubbelneer} & \textbf {\small 0.738 {\tiny( 0.07)}} \hphantom{\tiny \dubbelneer} \\
NP2004 & {\small 0.711 {\tiny( 0.05)}} & {\small 0.683 {\tiny( 0.07)}} {\tiny \dubbelneer} & {\small 0.688 {\tiny( 0.07)}} {\tiny \dubbelneer} & {\small 0.714 {\tiny( 0.05)}} \hphantom{\tiny \dubbelneer} & \textbf {\small 0.717 {\tiny( 0.06)}} \hphantom{\tiny \dubbelneer} \\
TD2004 & \textbf {\small 0.299 {\tiny( 0.04)}} & {\small 0.221 {\tiny( 0.03)}} {\tiny \dubbelneer} & {\small 0.220 {\tiny( 0.03)}} {\tiny \dubbelneer} & {\small 0.289 {\tiny( 0.03)}} {\tiny \enkelneer} & {\small 0.292 {\tiny( 0.03)}} \hphantom{\tiny \dubbelneer} \\
MQ2007 & {\small 0.340 {\tiny( 0.02)}} & \textbf {\small 0.370 {\tiny( 0.06)}} {\tiny \dubbelop} & {\small 0.352 {\tiny( 0.07)}} \hphantom{\tiny \dubbelneer} & {\small 0.336 {\tiny( 0.05)}} \hphantom{\tiny \dubbelneer} & {\small 0.339 {\tiny( 0.05)}} \hphantom{\tiny \dubbelneer} \\
MQ2008 & \textbf {\small 0.456 {\tiny( 0.05)}} & {\small 0.440 {\tiny( 0.10)}} \hphantom{\tiny \dubbelneer} & {\small 0.441 {\tiny( 0.08)}} \hphantom{\tiny \dubbelneer} & {\small 0.424 {\tiny( 0.10)}} {\tiny \dubbelneer} & {\small 0.426 {\tiny( 0.09)}} {\tiny \dubbelneer} \\
MSLR-WEB10k & \textbf {\small 0.301 {\tiny( 0.01)}} & {\small 0.292 {\tiny( 0.04)}} {\tiny \dubbelneer} & {\small 0.293 {\tiny( 0.04)}} {\tiny \enkelneer} & {\small 0.292 {\tiny( 0.04)}} {\tiny \dubbelneer} & {\small 0.293 {\tiny( 0.04)}} {\tiny \enkelneer} \\
OHSUMED & \textbf {\small 0.433 {\tiny( 0.05)}} & {\small 0.402 {\tiny( 0.06)}} {\tiny \dubbelneer} & {\small 0.404 {\tiny( 0.06)}} {\tiny \dubbelneer} & {\small 0.424 {\tiny( 0.06)}} \hphantom{\tiny \dubbelneer} & {\small 0.426 {\tiny( 0.06)}} \hphantom{\tiny \dubbelneer} \\
Yahoo & \textbf {\small 0.618 {\tiny( 0.05)}} & {\small 0.590 {\tiny( 0.08)}} {\tiny \dubbelneer} & {\small 0.600 {\tiny( 0.08)}} {\tiny \enkelneer} & {\small 0.616 {\tiny( 0.07)}} \hphantom{\tiny \dubbelneer} & {\small 0.610 {\tiny( 0.07)}} \hphantom{\tiny \dubbelneer} \\
\bottomrule
\end{tabular*}

\label{tab:offline}
\end{table*}

This section presents the results of our experiments and answers the research questions posed in Section~\ref{sec:intro}.

\subsection{Improving the user experience with \ac{Sim-MGD}}
First we consider \ref{rq:simmgd}: whether \ac{Sim-MGD} improves the user experience compared to \ac{MGD}. 

\subsubsection{Online performance.}
Table~\ref{tab:online} (Columns 2--4) displays the online performance of \ac{Sim-MGD} and \ac{MGD}. In the large majority of cases \ac{Sim-MGD} provides a significant increase in online performance over \ac{MGD}, both with the uniform and k-means document selection strategies. E.g., under the perfect user model, 7 out of 11 datasets for uniform and 8 out of 11 for k-means.  Significant decreases in online performance are found for \emph{HP2003}, \emph{TD2003}, \emph{TD2004} and \emph{OHSUMED} for uniform and for
\emph{TD2003}, \emph{TD2004} and \emph{OHSUMED} for k-means. Interestingly, all of these datasets model informational tasks, which suggests that it is more difficult to create an appropriate reference set in these cases. Furthermore, the differences between \ac{Sim-MGD} and \ac{MGD} are consistent over the different click-models. Therefore, we conclude that \ac{Sim-MGD} is as robust to noise as \ac{MGD}. 

Finally, Table~\ref{tab:online} (Columns 3--4) allows us to contrast the online performance of different document selection strategies: k-means beats uniform on the majority of datasets under all user models, and the noisier the user model is, the  bigger the majority is. Therefore, it seems that clustering results in a faster learning speed of \ac{Sim-MGD}; this could be because k-means will provide more dissimilar reference documents. Hence, the parameters in \ac{Sim-MGD} will be less correlated making learning faster than for uniform sampling.

\begin{figure}[tb]
\centering
\includegraphics[width=\columnwidth]{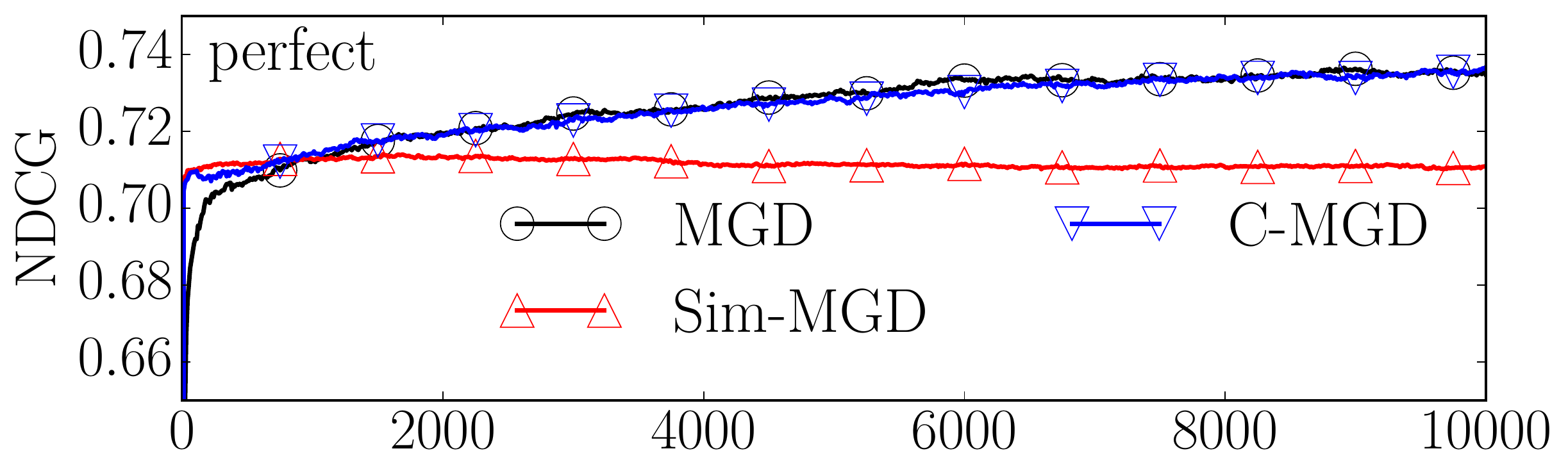}
\includegraphics[width=\columnwidth]{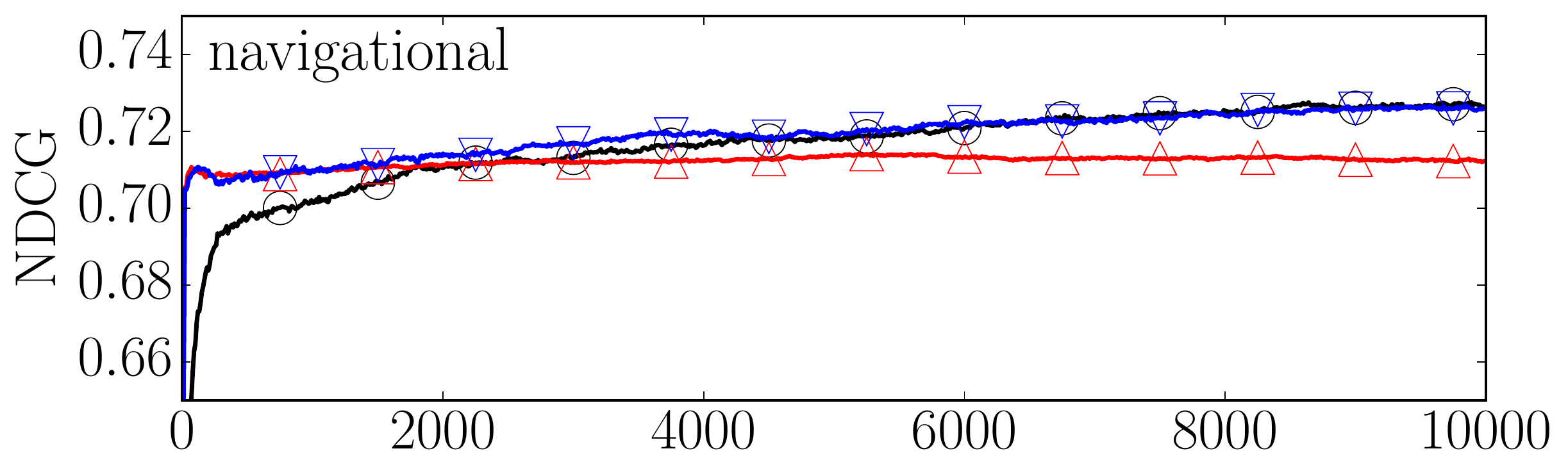}
\includegraphics[width=\columnwidth]{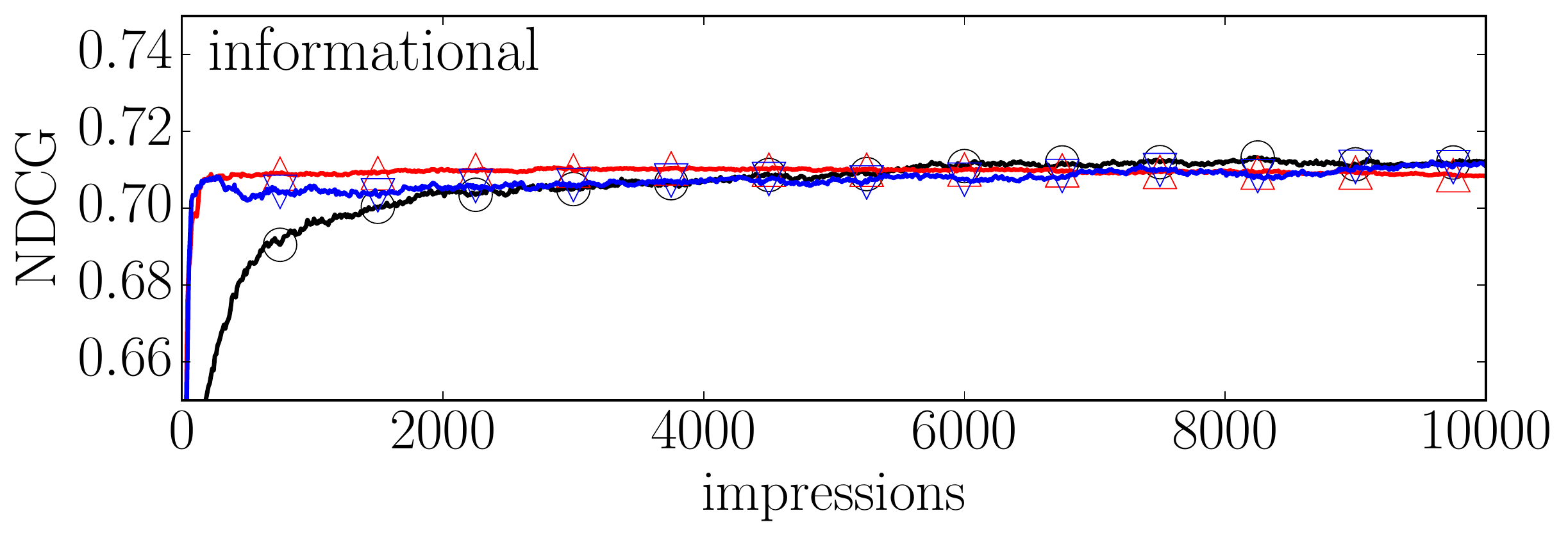}
\caption{Offline performance (NDCG) of \ac{MGD}, the \acs{Sim-MGD} and \ac{C-MGD} (k-means initialization) on the NP2003 dataset under three click models.}
\label{fig:mainoffline}
\end{figure}

In conclusion, \ac{Sim-MGD} improves the user experience most of the time, but is not reliable as it may provide a significantly worse experience depending on the dataset. 

\subsubsection{Offline performance.}
Table~\ref{tab:offline} (Columns 2--4) displays the offline performance of \ac{Sim-MGD} and \ac{MGD}. As predicted by the speed-quality tradeoff, we see that the convergence of \ac{Sim-MGD} after 10,000 impressions is substantially worse than \ac{MGD}. This suggests that the optimum found by \ac{MGD} can generally not be expressed by the similarity model in \ac{Sim-MGD}, i.e., it is not a linear combination of document features. 

Figure~\ref{fig:mainoffline} shows the offline performance of \ac{MGD} and \acs{Sim-MGD} on the \emph{NP2003} dataset for the three click models. Here, the improved learning speed is visible as \ac{Sim-MGD} outperforms \ac{MGD} in the initial phase of learning, under more click-noise \ac{MGD} requires more impressions to reach the same performance. For the \emph{informational} click model over 2000 impressions are required for \ac{MGD} to reach the performance \ac{Sim-MGD} had in fewer than 200. However, it is clear that \ac{Sim-MGD} has an inferior point of convergence, as it is eventually overtaken by \ac{MGD} under all click models. 

Lastly, Table~\ref{tab:offline} (Columns 3--4) shows the scores for \ac{Sim-MGD} with different reference document selection methods. The k-means selection method provides a higher online performance and a slightly better point of convergence. Therefore, it seems that clustering helps in selecting reference documents but has a limited effect.

To answer \ref{rq:simmgd}, \ac{Sim-MGD} improves the user experience in most cases, i.e., on most datasets and under all user models, with a consistent benefit for the k-means document selection strategy. As predicted by the speed-quality tradeoff, \ac{Sim-MGD} converges towards inferior rankings than \ac{MGD}, due to its less expressive model.

\subsection{Resolving the speed-quality tradeoff with \ac{C-MGD}}
Next, we address the speed-quality tradeoff with \ref{rq:cmgd}: whether \ac{C-MGD} is capable of improving the user experience while maintaining the state-of-the-art convergence point.

\subsubsection{Learning speed} To evaluate the user experience, the online performance of \ac{C-MGD} and \ac{MGD} can be examined in Table~\ref{tab:online} (Column~2 vs.~5 and~6). The online performance of \ac{C-MGD} is predominantly a significant improvement over that of \ac{MGD}. Moreover, when k-means document selection is used, no significant decreases are measured on any dataset or click model. Even on datasets where \ac{Sim-MGD} performs significantly worse than \ac{MGD} in terms of online performance, no significant decrease is observed for \ac{C-MGD}. Thus, \ac{C-MGD} deals with the inferior performance of its starting model by effectively switching to a more expressive model space.

\subsubsection{Quality convergence}
Furthermore, the quality side of the tradeoff is examined by considering the offline performance after 10,000 impressions, displayed in Table~\ref{tab:offline} (Column~2 vs.~5 and~6).
In the vast majority of cases \ac{C-MGD} shows no significant change in offline performance compared to \ac{MGD}.
For \acs{C-MGD} with uniform selection only four instances of significant decreases in offline performance w.r.t.\ \acs{MGD} are found scattered over different datasets and user models; this number is further reduced when k-mean selection is used. Only for MQ2008 under the informational user model this difference is greater than 0.1 NDCG. In all other cases, the offline performance of \acs{MGD} is maintained by \acs{C-MGD} or slightly improved. Conclusively, \acs{C-MGD} converges towards rankings of the same quality as \acs{MGD}.

\subsubsection{Switching models}
Lastly, we consider whether \ac{C-MGD} is able to effectively switch between model spaces. As discussed in the previous paragraphs, Table~\ref{tab:online} and~\ref{tab:offline} show that \ac{C-MGD} improves the user experience of \acs{MGD} while maintaining the final performance at convergence. This switching of models can also be observed in Figure~\ref{fig:mainoffline}, where the offline performance of \ac{Sim-MGD}, \ac{MGD} and \acs{Sim-MGD} on the \emph{NP2003} dataset for the three click models is displayed. As expected, we see that initially \acs{Sim-MGD} learns very fast and converges in less than 300 impressions; \ac{C-MGD} has the same performance during this period. When convergence of \acs{Sim-MGD} is approached \ac{C-MGD} switches to the linear model. A small drop in NDCG is visible when this happens under the informational click model. However, from this point on \ac{C-MGD} uses the same model as \ac{MGD} and eventually reaches a higher performance than it had before the switch was made. This indicates that the switch was made effectively but had some minor short-term costs, which can be accounted to the change in confidence: after switching, \ac{C-MGD} will perform more exploration in the new model space. As a result, \ac{C-MGD} may explore inferior parts of the model space before oscillating towards the optimum. Despite these costs, when switching \acs{C-MGD} is able to provide a reliable improvement in user experience over \ac{MGD} while \ac{Sim-MGD} cannot (Table~\ref{tab:online}). Thus we conclude that the switching of models is done effectively by \ac{C-MGD} as evident by the reliable improvement of online performance over \acs{MGD} while also having the same final offline performance.

In conclusion, we answer \ref{rq:cmgd} positively: our results show that in spite of the speed-quality tradeoff, \acs{C-MGD} improves the user experience of \acs{MGD} while still converging towards the same quality rankings. These findings were made across eleven datasets and varying levels of  noise in the user models employed.


\section{Conclusion}
\label{sec:conclusion}

In this paper we have addressed the speed-quality tradeoff that has been facing the field of \ac{OLTR}. Expressive models are capable of learning the most optimal rankings but require more user interactions and as a result frustrate more users during training. To put it bluntly, users may be frustrated in the initial phase of learning; models that converge at the best rankings frustrate the users for the longest period. 

As a solution we have introduced two methods. The first method is a ranking model that ranks by feature similarities with reference documents (\acs{Sim-MGD}). \acs{Sim-MGD} learns faster and consequently provides a much better initial user experience. As predicted by the speed-quality tradeoff it converges towards rankings inferior to \acs{MGD}. The second is a cascading approach, \acs{C-MGD}, that deals with the speed-quality tradeoff by using a cascade of models. Initially the simplest model in the cascade interacts with the users until convergence is detected; at this point a more expressive model continues the learning process. By doing so the cascade combines the best of both models: fast initial learning speed and optimal convergence.

The introduction of \acs{C-MGD} opens an array of possibilities.
A natural extension is to consider expressive models that have been successful in Offline-\acs{LTR} and place them in \acs{C-MGD} as the short-term user experience can be addressed by \ac{C-MGD}. E.g., an \acs{OLTR} version of LambdaMart \cite{burges2010ranknet} could be appended to
a cascade that starts with the \acs{Sim-MGD} model, then switches to \acs{MGD} and finally switches to a novel \acs{OLTR} regression forest.
Currently, there is no \acs{OLTR} method of gradient estimation for non-linear structures like regression trees: the introduction of \acs{C-MGD} removes an important hurdle for research into such methods. Additionally, an initialization method has to be introduced to enable the switch between such models. Ideally, the cascading approach should be extended to predict whether switching model space will have a positive effect; multileaving may be adapted to infer such differences.

\begin{spacing}{1}
\medskip\noindent\small
\textbf{Acknowledgments.}
This research was supported by
Ahold Delhaize,
Amsterdam Data Science,
the Bloomberg Research Grant program,
the Criteo Faculty Research Award program,
the Dutch national program COMMIT,
Elsevier,
the European Community's Seventh Framework Programme (FP7/2007-2013) under
grant agreement nr 312827 (VOX-Pol),
the Microsoft Research Ph.D.\ program,
the Netherlands Institute for Sound and Vision,
the Netherlands Organisation for Scientific Research (NWO)
under pro\-ject nrs
612.\-001.\-116, 
HOR-11-10, 
CI-14-25, 
652.\-002.\-001, 
612.\-001.\-551, 
652.\-001.\-003, 
and
Yandex.
All content represents the opinion of the authors, which is not necessarily shared or endorsed by their respective employers and/or sponsors.
\end{spacing}

\vspace*{-.5\baselineskip}

\balance

\bibliographystyle{abbrvnat}
 \bibliography{cikm2017online-similarity-ranking}

\end{document}